\documentclass[12pt,ams,preprint,a4paper]{revtex4}
\usepackage{graphicx}
\usepackage{amsmath,amssymb}

\begin{document}
\sloppy

\title{AN APPROACH TO THE COMPUTATION OF FEW/MANY--BODY MULTICHANNEL REACTIONS\footnote{\em Proceedings of the International Conference  `Nuclear Theory in the Supercomputing Era~--- 2016' \mbox{(NTSE-2016)}, Khabarovsk,
Russia, September 19--23, 2016. Eds.~A.~M.~\mbox{Shirokov} and  A.~I.~Mazur.
Pacific National University, Khabarovsk, Russia.}}

\newcommand{\Kurchatov}{National Research Center  "Kurchatov Institute",  123182 Moscow,  Russia}
\newcommand{\mifi}{National Research Nuclear University MEPhI, Moscow, Russia}
\author{V.~D.~Efros}\email{v.efros@mererand.com}\affiliation{\Kurchatov \\\mifi}


\begin{abstract}
A method to calculate reactions in quantum mechanics is outlined. It is advantageous, in particular, 
in problems with many open channels of various nature i.e. when energy is not low. 
In this method
there is no need  to specify reaction channels in a dynamics calculation. 
These channels come into play at merely the kinematics level and only
 after a dynamics calculation is done.
This calculation is of the
bound--state type while  continuum spectrum states never enter the game.
\end{abstract}

\bigskip


\maketitle

 

\section{Overview}

The approach reviewed in the paper is advantageous, in particular, 
in problems with many open channels of various nature i.e. when energy is not low. 
Conventional approaches dealing with continuum wave functions are impractical in such problems at least
at A$>3$.
The approach was successfully applied 
in nuclear reaction problems with $3\le$A$\le7$ and also recently for A=12 and 16 
proceeding from NN or NN+NNN forces. 
Many cases of reactions  induced by a
perturbation, i.e. electromagnetic or weak interaction,
were considered. Both inclusive (mostly) and exclusive processes were studied.
Reactions induced by strong interaction  still were not considered although 
this can be done in a 
similar way, see below.

The  main features of the approach
are the following. In a dynamics calculation in its framework
there is no need  to specify reaction channels {\it at all}. 
These come into play at merely the kinematics level and only
 after a dynamics calculation is done. Such a calculation 
is of the
bound--state type.

Correspondingly, continuum spectrum 
states never enter the game. In place of them, "response--like" functions, 
of the type of Eq. (\ref{r}) below, are 
basic ingredients of the approach. Reaction observables are expressed in terms of these
functions
 as quadratures, see Eqs. (\ref{t}) -- (\ref{r1}) below. It should  also be 
noted that in some problems of importance the quantities of Eq. (\ref{r}) form are of interest 
themselves representing observable response 
functions for inclusive perturbation--induced reactions.

And the required "response--like" functions of Eq. (\ref{r}) form are 
obtained not in terms of  the complicated continuum spectrum 
states  entering their definition but via a 
bound--state type calculation. As the first step, an integral transform of such a
function
is performed. The transform is found in a closed form and 
represents a "continuum sum rule" depending on a $\sigma$ parameter,
Eq. (\ref{inteq}). It is evaluated via a bound--state type calculation. As the next step, this sum rule
is considered as an equation determining the "response--like" function, i.e. its inversion is performed.
Once this is done, the above mentioned quadratures giving the reaction observables are readily
obtained.

Thus, as claimed above, the specification of reaction channels in the dynamics calculation
and dealing with continuum wave functions are avoided in this approach. The criterion of accuracy 
is stability of response--like function obtained. 

In addition to checking the stability,
 comparisons with more conventional calculations that deal with continuum wave 
functions have been performed. In the benchmark paper \cite{gloeckle} the Faddeev results for
the $^3$H photoabsorption total cross section are compared with the results \cite{np} obtained via
the above described approach. In the framework of this approach, the Lorentz integral transform (LIT), 
see the next section, was used. The Argonne V18 NN interaction and 
the Urbana IX NNN potential have been employed. The results are shown in Fig. 1.

A compete agreement is observed in the case when only the NN force is retained 
while in the case when the NNN force is added such an agreement 
is observed everywhere except for the peak region where a slight difference is present. 
In Ref. \cite{EIHH} the LIT results in the same problem have been obtained employing expansions over
two different bases at solving the dynamics equation, the correlated hyperspherical 
basis (CHH) and the effective interaction 
hyperspherical basis (EIHH). The results are shown in Fig. 2.
 
\begin{figure}
\centerline{\includegraphics[width=0.4\textwidth]{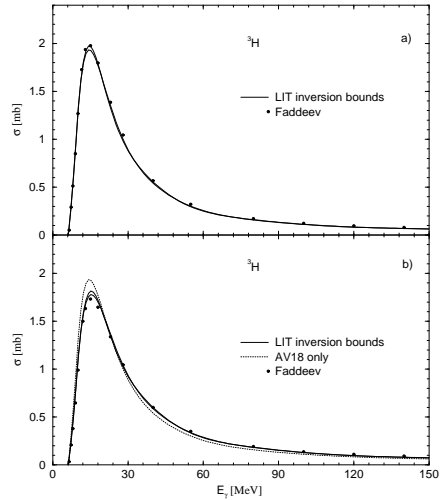}}
\caption{Comparison of the Faddeev and LIT results for the total 
$^3$H photoabsorption cross section in unretarded dipole approximation
(a) with NN (AV18) force  only and (b) with NN(AV18)+NNN(UrbIX) force. The dots are the Faddeev results
and the two curves represent the bounds for the inversion of the LIT.
The dotted curve in (b) is the result with AV18 only.}
\label{fig1}      
\end{figure}

\begin{figure}
\centerline{\includegraphics[width=0.4\textwidth]{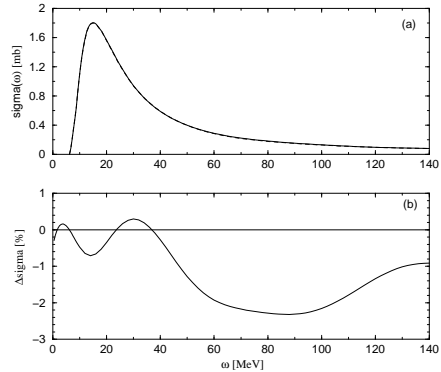}}
\caption{(a) The same cross section with the same NN+NNN force as 
in Fig.~1(b) ($E_\gamma\rightarrow\omega$). 
It is obtained with the help of LIT
at solving the dynamics equation in two ways. The full curve and the dashed
curve represent its solution using the CHH and EIHH expansions, respectively. (b) The 
relative difference \mbox{$[\sigma({\rm CHH})-\sigma({\rm EIHH})]/\sigma({\rm CHH})$} 
between the two solutions.}
\label{fig2}      
\end{figure}

These results practically coincide with each other which 
testifies to that the LIT results in Fig. 1 b) are 
accurate.

\begin{figure}
\centerline{\includegraphics[width=0.4\textwidth]{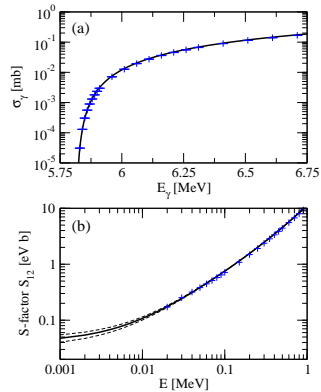}}
\caption{(a) The  total 
$^3$He photoabsorption cross section in the threshold region calculated with the MT NN potential. 
The full curve represents the LIT results and the plus signs represent the results from the direct calculation with explicit continuum
wave functions; (b) same results as in (a) but rescaled in order to determine the \mbox{$S$--factor}.  
The inversion error bounds are shown by dashed lines. E denotes the p--d relative motion energy}
\label{fig3}      
\end{figure}

In Fig. 3 one more test is presented \cite{wl}. 
The total cross section of the $^3$He($\gamma$,p)d reaction
in the threshold region is calculated in two ways, from the LIT, as above, and via a 
direct calculation of the pd continuum wave functions.  In this case, there is no real need for use
of the  method of integral transforms since the problem is a one--channel one. Another point is that the
problem considered is unfavorable for this method since the cross section at the threshold is tiny and  
the values of the response function at such energies 
contribute extremely little to the integral 
pertaining to 
the corresponding 
integral equation of Eq. (\ref{inteq}) and therefore to the input to solve the equation.   
Despite this, a 
complete agreement of the results of the two methods is observed. This is most 
clearly seen from Fig. 3 (b)
where the quickly varying Gamow factor is factored out 
from the cross section,  and the remaining astrophysical 
$S$--factor  is presented.  The central Malfliet--Tjon NN
potential was employed in this calculation. Let us also mention that the pd continuum 
wave functions that have led to the results in Fig. 3 provide phase shifts practically coinciding 
with those of the other group \cite{Kie16}.   
 
In  the next section basic points of the approach are presented, and Sec. 3 contains further comments.

\section{Basics of the Method}

Let us consider "response--like" quantities 
having the structure  
\begin{equation} R(E)=\sum_n\langle Q'|\Psi_n\rangle\langle\Psi_n|Q\rangle\delta(E-E_n)+
\sum\!\!\!\!\!\!\!\int d\gamma\langle Q'|\Psi_\gamma\rangle\langle
\Psi_\gamma|Q\rangle\delta(E-E_\gamma).\label{r}\end{equation}
Here $\Psi_n$ and $\Psi_\gamma$ denote, respectively, bound states and continuum spectrum states with energies $E_n$ and
$E_\gamma$
pertaining to the Hamiltonian of a problem. 
The $\gamma$ subscript labeling the states may include both continuous and discrete 
variables which is reflected in the sum over integral symbol. The set of states is complete which
may be written in the form
\begin{equation} \sum_n|\Psi_n\rangle\langle\Psi_n|+
\sum\!\!\!\!\!\!\!\int d\gamma|\Psi_\gamma\rangle\langle\Psi_\gamma|=I\label{i}\end{equation}
where $I$ is the identity operator. This implies the 
normalizations \mbox{$\langle\Psi_n|\Psi_{n'}\rangle=\delta_{n,n'}$} and
\mbox{$\langle\Psi_\gamma|\Psi_{\gamma'}\rangle=\delta(\gamma-\gamma')$}. 
As said above, in the present approach
reaction observables are expressed   in terms of quantities (\ref{r}) as quadratures.
First we shall discuss this point. After that,
the above mentioned 
evaluation of quantities (\ref{r}) via bound--state methods will be outlined.

 Consider the case of a reaction induced by strong interaction.
Let $\phi_{i}(E)$ be the product of bound states of fragments and of a
factor representing their free relative motion in the initial state. Let 
$\phi_{f}(E)$ be such products pertaining to final states. Let us 
denote ${\cal A}$ the operator of antisymmetrization with ${\cal A}^2={\cal A}$. So that  
\mbox{${\cal A}\phi_i(E)$} and \mbox{${\cal A}\phi_f(E)$} are 
the antisymmetrized "free--motion" states \cite{GW}.
Let us use the notation \mbox{${\bar\phi}_i(E)={\cal A}(H-E)\phi_i(E)$} and
\mbox{$ {\bar\phi}_f(E)={\cal A}(H-E)\phi_f(E)$} where $H$ is the Hamiltonian.
This may be written as  \mbox{$\bar\phi_{i}={\cal A}V_{i}^{res}\phi_{i}$} and 
$\bar\phi_{f}={\cal A}V_{f}^{res}\phi_{f}$ where 
$V_{i,f}^{res}$ are interactions between fragments pertaining to a channel.
These interactions are assumed here to be of a short range.
The point on long--range inter--fragment Coulomb interactions
is commented below.

The reaction $T$ matrix is  \cite{GW}
\begin{equation} T_{fi}=T_{fi}^{Born}+
\langle{\bar\phi}_f(E)|(E-H+i\epsilon)^{-1}|{\bar\phi}_i(E)\rangle,\label{t}\end{equation}
\mbox{$\epsilon\rightarrow+0$}. Here $T_{fi}^{Born}$ denotes the  Born contribution,
$$
T_{fi}^{Born}=\langle\phi_f|{\bar\phi}_i\rangle=\langle{\bar\phi}_f|\phi_i\rangle.
$$
The problem lies in calculating the non--Born contribution involving
the Green function \mbox{$(E-H+i\epsilon)^{-1}$}. 
Let us introduce the quantity
\begin{eqnarray} R_E(E')=
\sum_n \langle {\bar\phi}_f(E)|\Psi_n\rangle\langle\Psi_n|{\bar\phi}_i(E)\rangle
\delta(E'-E_n)\nonumber\\
+\sum\!\!\!\!\!\!\!\int d\gamma
\langle {\bar\phi}_f(E)|\Psi_\gamma\rangle\langle\Psi_\gamma|{\bar\phi}_i(E)\rangle
\delta(E'-E_\gamma).\label{resp}\end{eqnarray}
This quantity has the same structure as that in Eq.~(\ref{r}) (with the replacement \mbox{$E\rightarrow E'$}).
And the contribution to the $T$ matrix we discuss may  readily be calculated in its
terms as
\begin{equation}\int dE' R_E(E')(E-E'+i\epsilon)^{-1}\equiv-i\pi R_E(E)+P\int dE' R_E(E')(E-E')^{-1}.
\label{rint}\end{equation}
Thus, indeed, reaction cross sections
may be expressed in terms of the "response--like" quantities
of Eq. (\ref{r}) as quadratures.

The amplitude of a perturbation--induced reaction is \mbox{$\langle\Psi^-_f|\hat{O}|\Psi_0\rangle$}
 where  
${\hat O}$ is a perturbation, $\Psi_0$ is an unperturbed initial bound state, and $\Psi^-_f$ is 
a continuum spectrum state. To calculate this amplitude let us substitute  the expression
\cite{GW}
$$
\langle\Psi^-_f|=\langle\phi_f|+\langle{\bar\phi_f}|(E-H+i\epsilon)^{-1}
$$
in it. Then 
\begin{equation}\langle\Psi^-_f|O|\Psi_0\rangle=\langle\phi_f|O|\Psi_0\rangle+
\langle{\bar \phi}_f|(E-H+i\epsilon)^{-1}|O\Psi_0\rangle\label{r1}\end{equation}
and one may proceed as above 
with  the replacement \mbox{${\bar\phi}_i\rightarrow {\hat O}\Psi_0$}
there. 

A modification of the above relations required to incorporate the
long--range inter--fragment Coulomb interactions is outlined in Ref. \cite{efr99}.
(And if the response function itself is the objective of a calculation then the Coulomb interaction 
requires no special consideration as
seen from below.)
This modification leads to modified $Q$ and $Q'$ states which
include Coulomb functions in the inner region of the relative motion of fragments.
Of course, it is very easy to obtain such Coulomb functions in the case of two--fragment reaction
channels.  

Now we need to consider the calculation of quantities having the "response--like" structure of
Eq.~(\ref{r}). We write them as   
\begin{equation} R(E)=\sum_n R_n\delta(E-E_n)+f(E),\qquad R_n=\langle Q'|\Psi_n\rangle 
\langle\Psi_n|Q\rangle,\label{c2}\end{equation}
\begin{equation} f(E)=
\sum\!\!\!\!\!\!\!\int d\gamma\langle Q'|\Psi_\gamma\rangle\langle\Psi_\gamma|Q\rangle
\delta(E-E_\gamma)\label{f}.\end{equation}
The main problem consists in calculating the contribution (\ref{f}).
And this should be done avoiding the calculation of multiparticle continuum states entering it.

First let us list the simple sum--rule result. With the help of Eq.~(\ref{i}) 
one obtains
\begin{equation}\int_{E_{thr}}^\infty f(E)dE+\sum_n R_n=\langle Q'|Q\rangle.\label{s}\end{equation}
Here $E_{thr}$ is the continuum spectrum threshold value so that $f(E)$ varies in the range 
\mbox{$E_{thr}\le E\le\infty$}. While this single sum rule does not determine $R(E)$, this
goal can be achieved with the help of "generalized" sums of the form
\begin{equation} \int K(\sigma,E)R(E)dE\label{ineq}\end{equation}
depending on a continuous parameter.
They  are equal to
\begin{eqnarray}
\sum\!\!\!\!\!\!\!\int d\gamma\langle Q'|\Psi_\gamma\rangle K(\sigma,E_\gamma)
\langle\Psi_\gamma|Q\rangle+\sum_n\langle Q'|\Psi_n\rangle K(\sigma,E_n)
\langle\Psi_n|Q\rangle.
\end{eqnarray}
Taking into account Eq. (\ref{i}) this quantity can be represented as
\mbox{$\langle Q'|K(\sigma,H)|Q\rangle$}
where, as above, $H$ is the Hamiltonian of a problem. 
Thus, one 
comes to the  relation
\begin{equation}
\int_{E_{thr}}^\infty K(\sigma,E)f(E)dE+\sum_nK(\sigma,E_n)R_n=\Phi(\sigma),\qquad 
\Phi(\sigma)\equiv\langle Q'|K(\sigma,H)|Q\rangle
\label{inteq}
\end{equation}
where $f(E)$ and $R_n$ are the continuous part of the response--like function  $R(E)$ 
and discrete contributions to it, see Eqs. (\ref{f}) and (\ref{c2}).
Since this relation is valid for any~$\sigma$ it may be considered as an equation to determine $R(E)$,
i.e. $f(E)$ and $R_n$,
 provided that  one is able to calculate 
the quantity \mbox{$\langle Q'|K(\sigma,H)|Q\rangle$}. At many $K$ kernels this equation determines
$f(E)$ and $R_n$ in a unique way.

\section{Further comments}

Thus the equation of Eq. (\ref{inteq}) form is to be solved. First, one needs to calculate its
right--hand side input. If one is able to diagonalize the Hamiltonian on a sufficiently big subspace 
of basis functions this can be readily done. In this case, one can use the approximation of the type
\begin{equation}
\langle Q'|K(\sigma,H)|Q\rangle\simeq\sum_{n=1}^N\langle Q'|\varphi_n^N\rangle 
K(\sigma,E_n^N)\langle\varphi_n^N|Q\rangle.\label{discr}
\end{equation}
Here $N$ is the dimension of the subspace and other notation is obvious. 
Suppose, for example, that the  
$K(\sigma,E)=\exp[-(\sigma-E)^2/\sigma_0^2]$ kernel is employed. At a given 
accuracy in the input $\Phi(\sigma)$,  smaller $\sigma_0$
values would lead to a better reproduction of details of $f(E)$ at solving Eq. (\ref{inteq}).
Indeed, at large $\sigma_0$ values,
contributions to $\Phi(\sigma)$ from peculiarities of $f(E)$ are spread over large $\sigma$ intervals,
and sizes of these contributions 
may be comparable with sizes of inaccuracies in calculated $\Phi(\sigma)$.  
At the same time, smaller $\sigma_0$ values require use of subspaces of basis functions of
higher dimension. Indeed, accurate $\Phi(\sigma)$ 
values emerge only at such sizes of these subspaces  that (at $\sigma$ values
of significance) energy ranges $\sigma-\sigma_0\le E\le\sigma+\sigma_0$ contain
sufficiently many $E_n^N$ eigenvalues.

The right--hand side of Eq. (\ref{discr}) represents the result of smoothing the pseudo--response
$$
\sum_{n=1}^N\langle Q'|\varphi_n^N\rangle \langle\varphi_n^N|Q\rangle\delta(E-E_n^N)
$$  
with the help of the smoothing function $K(\sigma,E)$. Such type smoothings were performed in the
literature and their results were considered as approximations to true responses for 
perturbation--induced   inclusive reactions. In the difference to this, in the present approach 
such results are not adopted to be approximations to true responses. Here they play the role of
the input to the integral equation and final true responses emerge as its solution. This 
refinement makes possible
to obtain consistent and more accurate results.

At some choices of the kernel $K$ it is possible to calculate the input $\Phi(\sigma)$ to 
Eq.~(\ref{inteq}) also 
without the diagonalization of the Hamiltonian. The simplest case is the Stieltjes kernel
$K(\sigma,E)=(\sigma+E)^{-1}$ where $\sigma$ is real and lies apart from the spectrum of a Hamiltonian.
 In this case one has 
\begin{equation}
\Phi(\sigma)=\langle Q'|{\tilde \psi}\rangle,\qquad {\tilde \psi}=(H+\sigma)^{-1}Q.
\end{equation}
I.e. ${\tilde \psi}$ is the solution to the inhomogeneous Schr\"odinger--like equation
\begin{equation}
(H+\sigma){\tilde \psi}=Q.\label{seq}
\end{equation}
From the fact that $\langle Q|Q\rangle$ is finite it follows that the solution
is localized, and such a solution is unique. Another case is the so called Lorentz kernel
\begin{equation} 
K(\sigma=\sigma_R+i\sigma_I,E)=1/[(\sigma_R-E)^2+\sigma_I^2].
\end{equation}
Writing 
\begin{equation}
\frac{1}{(\sigma_R-E)^2+\sigma_I^2}=\frac{1}{2i\sigma_I}\left(\frac{1}{\sigma_R-E-i\sigma_I}-
\frac{1}{\sigma_R-E+i\sigma_I}\right)
\end{equation}
one reduces the calculation in this case to that in the Stiltjes one but with complex kernels.
The solutions to the correponding Eq. (\ref{seq}) type equations 
are localized and unique also in this case. 

Since the Lorentz kernel has a limited range,  inversion of the
 transform is more accurate in the Lorentz case than that in the Stiltjes case 
at the same accuracy in the input, c.f. the reasoning above. Still, when 
an expansion over a basis is applied to solve Eq. (\ref{seq}) type equations, convergence 
of $\Phi(\sigma)$
in the Stiltjes case  is faster  than that in the Lorentz case with a small $\sigma_I$.
Indeed, at $\sigma_I\rightarrow 0$ the continuum spectrum regime is recovered at $\sigma_R$ values
of interest belonging to the scattering line. 

One more case is the Laplace kernel $K(\sigma,E)=\exp(-\sigma E)$. The corresponding 
\mbox{$\langle Q'|e^{-\sigma H}|Q\rangle$} input
can be calculated with 
the Green Function Monte Carlo method. 

We shall not discuss here the point of solving Eq. (\ref{inteq}), i.e. the inversion of the transform,
referring for this to the literature. Let us mention only that such an equation
represents a classical "ill--posed problem". (This does not mean at all that the problem is a 
very difficult one!)
A standard 
regularization procedure was applied in practical calculations and convergent results
have been obtained. Still, with such a procedure 
a sufficient accuracy of the input $\Phi(\sigma)$ 
 may be harder to achieve in problems with not small number of particles. 
A new method to solve Eq. (\ref{inteq}) was proposed recently \cite{efr12}. 
In this method, the number of
maxima and minima of the solution sought for is imposed as an additional constraint. 
The method does not require a regularization. It has been proved 
that the method is convergent at least everywhere 
 except for the points of maxima and minima of $f(E)$. 
Thus, apart from this restriction, 
the problem becomes a well--posed one with the constraint imposed.
 With the same
approximate inputs the method provides far more accurate results than 
the standard regularization procedure in simple examples considered. But its further study 
is still required.

The discrete contributions $R_n$ in Eq. (\ref{inteq}) may be calculated separately. For a convenient way
to do this in the case of the Lorentz or Stiltjes transform see \cite{efr99}. Another option is 
the following. The general algorythm applied at solving Eq. (\ref{inteq}) consisted in writing an ansatz
for $f(E)$ that included parameters and in fitting the parameters to $\Phi(\sigma)$. 
The $R_n$ amplitudes and, if expedient, $E_n$ energies may be included in the set of such parameters.  

A limitation of the present approach is that in order to reproduce fine details of spectra of 
reactions, such as the widths of narrow resonances, an increased accuracy in the input 
$\Phi(\sigma)$ is required. The reason is the same as that discussed above in connection with Eq.~(\ref{discr}). 
This feature is similar to the situation with extracting widths of 
narrow resonances from experiments in which scattering or reaction cross sections are measured.
Still, narrow resonances are normally located at low energy whereas the present method is designed
for calculations of reactions when many channels are open, i.e. not at low energy. Then information
on the widths of narrow resonaces taken from experiment or from 
alternative calculations may be readily incorporated 
in an algorythm of solving Eq. (\ref{inteq}). And, anyway, inaccuracies
in widths of resonances at low energy in the present method would not lead to inaccuracies 
at reproducing reaction spectra at higher energy. In addition, in the light nuclei case 
widths of resonances are normally not so narrow. In Ref. \cite{leid} the width about
200 KeV of such 
a resonance in $^4$He was reproduced with a 
reasonable accuracy in the framework of the present approach.
See also Fig. 3 above in this regard.

In conclusion, the relevant literature is listed in addition to the references above. 
The approach to calculate reactions described in Sec. 2 has been introduced in Ref.~\cite{efr85}.
Its presentation here is close to Ref. \cite{efr99}.
The bound--state type, i.e. sum--rule, calculation of integral transforms  
of observable
responses $R(E)$, i.e. pertaining to inclusive perturbation--induced reactions,
   has been
 suggested in \cite{efr80} considering the Stieltjes transform 
and in \cite{lapl} in the Laplace transform case. No inversions of the transforms
were considered there. An alternative approach \cite{lang} was also developed in which
an observable
$R(E)$ is reconstructed from its moments of the type $\langle E^{-n}\rangle$, $n=0,\ldots,N$. 
The quantity of Eq. (\ref{s})
represents then the zero moment. Subsequent moments are calculated recursively. That approach 
referred only 
to the case of inclusive perturbation--induced reactions, in the difference to the above described method 
\cite{efr85} of treatment
of general--type reactions, i.e. exclusive  
perturbation-- and strong interaction--induced ones. 
The described way to calculate $\Phi(\sigma)$ involving Eq.~(\ref{discr}) was suggested
in Ref. \cite{efr99} (although at too restrictive conditions imposed on $Q$ and $Q'$). Prior to this  such
a calculation of the Lorentz transform in a particular problem has been performed by I.J. Thompson.
The Lorentz transform has been introduced in the present context in Ref. \cite{ELO94}.
Its evaluation in the form listed above was given in Ref. \cite{efr99}. 
In Ref. \cite{mar} an efficient
algorythm to calculate $\Phi(\sigma)$ at solving Eq. (\ref{seq}) with the help of 
an expansion over basis functions was developed. 
In the review papers \cite{efr99} and \cite{ELOB07} the subject of the transform inversion 
in the framework of the customary approach is
considered, in particular. In Ref. \cite{ELOB07}  earlier applications done
with the help of the Lorentz transform are reviewed as well. 
Among later applications,
advances in study of
heavier nuclei \cite{o16,c12} are to be mentioned. Existing bound--state techniques 
to solve the
multiparticle scattering problem  are reviewed in Ref. \cite{carb}.


\begin{thebibliography}{99}

\bibitem{gloeckle} J. Golak, W.
Gl\"ockle, H. Kamada, A. Nogga,
R. Skibi\'nski, H. Witala, V.D. Efros, W. Leidemann, G. Orlandini
and E.L. Tomusiak, Nucl. Phys. A {\bf 707},  365 (2002). 

\bibitem{np} V.D. Efros, W. Leidemann, 
G. Orlandini and E.L. Tomusiak, Nucl. Phys. A {\bf 689}, 421c (2001). 

\bibitem{EIHH} N. Barnea, W. Leidemann, G. Orlandini, V.D. Efros
and E.L. Tomusiak, Few--Body Systems {\bf 39}, 1 (2006).

\bibitem{wl} S. Deflorian,  V.D. Efros  and W. Leidemann, Few--Body Systems {\bf 58}: 3 (2017).

\bibitem{Kie16} A. Kievsky, priv. comm. (2016).

\bibitem{GW} M. Goldberger and K. Watson, {\it Collision Theory}, Dover publications, New York, 2004.
\bibitem{efr99} V.D. Efros, Yad. Fiz.  {\bf 62}, 1975 (1999)  [Phys. At. Nucl. 
{\bf 62},  1833 (1999)] (arXiv: \mbox{nucl-th/9903024}).
\bibitem{efr12} V.D. Efros, Phys. Rev. E {\bf 86},  016704 (2012).
\bibitem{leid} W. Leidemann, Phys. Rev. C {\bf 91}, 054001 (2015). 
\bibitem{efr85} V.D. Efros, Yad. Fiz. {\bf 41},  1498 (1985) [Sov. J. Nucl. Phys {\bf 41}, 949 (1985)].
\bibitem{efr80} V.D. Efros, Ukr. Fiz. Zh. {\bf 25}, 907 (1980)[Ukr. Phys. J.]. 
\bibitem{lapl} D. Thirumalai and B.J. Berne, J. Chem. Phys. {\bf 79}, 5029 (1983).
\bibitem{lang}W.P. Reinhardt, 
        {\it Theory of Applications of Moment Methods in
        Many--Fermion Systems}, ed.~B.J.~Dalton {\it et al}, Plenum, N.Y., 1980, p.~129;
        P.W.~Langhoff, {\it ibid}, p.~191.
\bibitem{mar} M.A. Marchisio, N. Barnea, W. Leidemann and G. Orlandini, Few--Body Systems {\bf 33}, 259
(2003).
\bibitem{ELO94}V.D. Efros, W. Leidemann and G. Orlandini, Phys. Lett. {\bf B338},
130 (1994).
\bibitem{ELOB07}V.D. Efros, W. Leidemann, G. Orlandini and N. Barnea, J. Phys. G: Nucl. Part. Phys. {\bf 34}, R459 (2007).
\bibitem{o16}S. Bacca, N. Barnea, G. Hagen, G. Orlandini and T. Papenbrock, Phys. Rev. Lett. {\bf 111},
122502 (2013).
\bibitem{c12}A. Lovato, S. Gandolfi, J. Carlson, S.C. Pieper and R. Schiavilla, Phys. Rev. Lett. 
{\bf 117}, 082501 (2016).
\bibitem{carb} J. Carbonell, A. Deltuva, A.C. Fonseca and R. Lazauskas, Progr. Part. Nucl. Phys. 
{\bf 74}, 55 (2014).
\end{thebibliography}
\end{document}